\documentclass[%
 reprint,
 superscriptaddress,
 amsmath,amssymb,
 aps,
 prd,
 floatfix,
]{revtex4-2}
\usepackage{float}
\usepackage{graphicx}
\usepackage{dcolumn}
\usepackage{bm}
\usepackage{xcolor}

\usepackage{hyperref}

\usepackage{soul}
\usepackage{comment}
\usepackage{siunitx}
\usepackage{orcidlink}

\usepackage[caption=false]{subfig}
\begin{document}


\title{Spiral Tuning of Wire-metamaterial Cavity for Plasma Haloscope}

\author{Jacob Lindahl}
\affiliation{The Oskar Klein Centre, Department of Physics, Stockholm University, AlbaNova, SE-10691 Stockholm, Sweden}

\author{Rustam Balafendiev}
\affiliation{Science Institute, University of Iceland, Dunhagi 5, 107, Reykjavik, Iceland}
\affiliation{School of Physics and Engineering, ITMO University, Kronverksky Pr. 49, 197101, St. Petersburg, Russia}

\author{Gagandeep Kaur}
\affiliation{The Oskar Klein Centre, Department of Physics, Stockholm University, AlbaNova, SE-10691 Stockholm, Sweden}

\author{Gaganpreet Singh}
\affiliation{The Oskar Klein Centre, Department of Physics, Stockholm University, AlbaNova, SE-10691 Stockholm, Sweden}

\author{Andrea Gallo Rosso}
\affiliation{The Oskar Klein Centre, Department of Physics, Stockholm University, AlbaNova, SE-10691 Stockholm, Sweden}

\author{Jan Conrad}
\affiliation{The Oskar Klein Centre, Department of Physics, Stockholm University, AlbaNova, SE-10691 Stockholm, Sweden}

\author{Jon E. Gudmundsson}
\affiliation{Science Institute, University of Iceland, Dunhagi 5, 107, Reykjavik, Iceland}
\affiliation{The Oskar Klein Centre, Department of Physics, Stockholm University, AlbaNova, SE-10691 Stockholm, Sweden}

\author{Junu Jeong\,\orcidlink{0000-0003-0194-9587}}
\email{jun-woo.jeoung@fysik.su.se}
\affiliation{The Oskar Klein Centre, Department of Physics, Stockholm University, AlbaNova, SE-10691 Stockholm, Sweden}

\date{\today}

\begin{abstract}
Axions are hypothetical particles that provide a compelling solution to two major mysteries in modern physics: the strong CP problem and the nature of dark matter.
The plasma haloscope has been proposed as a promising approach for probing the higher-mass regime for dark matter axions by employing a periodic arrangement of conducting wires.
In this work, we introduce a novel tuning mechanism for such wire-based structures by arranging the wires into a spiral configuration.
This design enables continuous frequency tuning of 25\% with a single central rotation while maintaining the form factor.
It also achieves scanning speeds several times faster than traditional tuning approaches, primarily due to the circular perimeter geometry, making it well suited for solenoidal magnet bores.
To validate the concept, we fabricated a prototype cavity with six spiral arms and experimentally demonstrated its feasibility, obtaining frequency tuning in close agreement with numerical simulations.
\end{abstract}

\maketitle

\section{Introduction}

Axions are hypothetical particles proposed by the Peccei–Quinn mechanism to dynamically resolve the charge-parity symmetry problem in quantum chromodynamics~\cite{PecceiQuinn:PRL:1977,Weinberg:PRL:1978,Wilczek:PRL:1978}.
Ultralight axions with masses below the electronvolt scale are strong dark matter candidates due to their non-relativistic production in the early Universe and extremely weak interactions with Standard Model particles~\cite{Preskill:PLB:1983,Abbott:PLB:1983,Dine:PLB:1983}.
Prominent theoretical models include the Kim–Shifman–Vainshtein–Zakharov (KSVZ)~\cite{Kim:PRL:1979,Shiftman:NPB:1980} and Dine–Fischler–Srednicki–Zhitnitsky (DFSZ)~\cite{Zhitnitsky:SJNP:1980,Dine:PLB:1981} frameworks.

Numerous experimental efforts worldwide are dedicated to the search for axion dark matter~\cite{IRASTORZA:PPNP:2018,AxionLimits,Semertzidis:SciAdv:2022}.
Among these, the cavity haloscope technique~\cite{Sikivie:PRL:1983} remains the most sensitive, as it has been reaching out to the theoretically predicted axion-photon coupling in the microwave frequency range.
However, the sensitivity of this method decreases at higher axion masses, primarily because the resonant volume diminishes with increasing frequency, thereby reducing the scanning speed.

Recent independent theoretical studies for early-universe simulations, based on the post-inflationary scenario, indicate a preference for axion masses above 10\,GHz to account for the observed dark matter abundance~\cite{Kawasaki:PRD:2015,Fleury:JCAP:2016,Gorghetto:SciPost:2021,Kim:JHEP:2024,Saikawa:JCAP:2024,Benabou:PRL:2025}.
This motivates the further development of experimental strategies capable of probing this high-frequency regime~\cite{Marsh:PRL:2019,Majorovits:JP:2020,Liu:PRL:2022,Chiles:PRL:2022,Bajjali:JCAP:2023,Jeong:PRD:2023,Miguel:PRD:2025}.

One promising approach is the plasma haloscope~\cite{Lawson:PRL:2019}, which employs a periodic array of conducting wires to form a metamaterial medium with a tunable effective plasma frequency.
This configuration supports efficient axion–photon conversion by matching the axion mass with the plasma frequency of the metamaterial structure.

The periodic wire array can be interpreted as a cavity resonator, where the fundamental transverse magnetic (TM) mode resonates near the effective plasma frequency.
The resonant frequency depends strongly on the spacing between adjacent wires, necessitating coordinated mechanical adjustments for tuning.
Several methods have been proposed to achieve a tunability of this design~\cite{Lawson:PRL:2019,Balafendiev:arXiv:2025,Goulart:arXiv:2025}.

In this work, we introduce a novel tuning method based on bundling wires into a spiral configuration.
This design enables frequency tuning through a single central rotation that simultaneously adjusts the spacing between adjacent spiral bundles.
To validate this concept, we designed and fabricated a prototype cavity and demonstrated that its measured resonant behavior is in agreement with numerical simulations.

Section~\ref{sec:spiral} introduces the underlying principles of the spiral tuning method, presents a design protocol for determining wire placement along the spiral arms, and evaluates its performance against alternative methods.
Section~\ref{sec:meas} describes the prototype cavity construction and measurement procedure, and compares the results with simulations.
In Section~\ref{sec:discussion}, we discuss the advantages of the spiral-arm geometry, possible optimizations, and its application to plasma haloscope experiments.

\section{\label{sec:spiral}Spiral Tuning Method}

A periodic array of conducting wires forms a wire metamaterial characterized by a reduced effective plasma frequency.
This frequency is predominantly determined by the wire diameter and spacing.
When the effective plasma frequency matches the axion mass, axion–photon conversion is resonantly enhanced, enabling sensitive searches in the high-mass regime.

\begin{figure}
    \centering
    \includegraphics[width=1.0\linewidth]{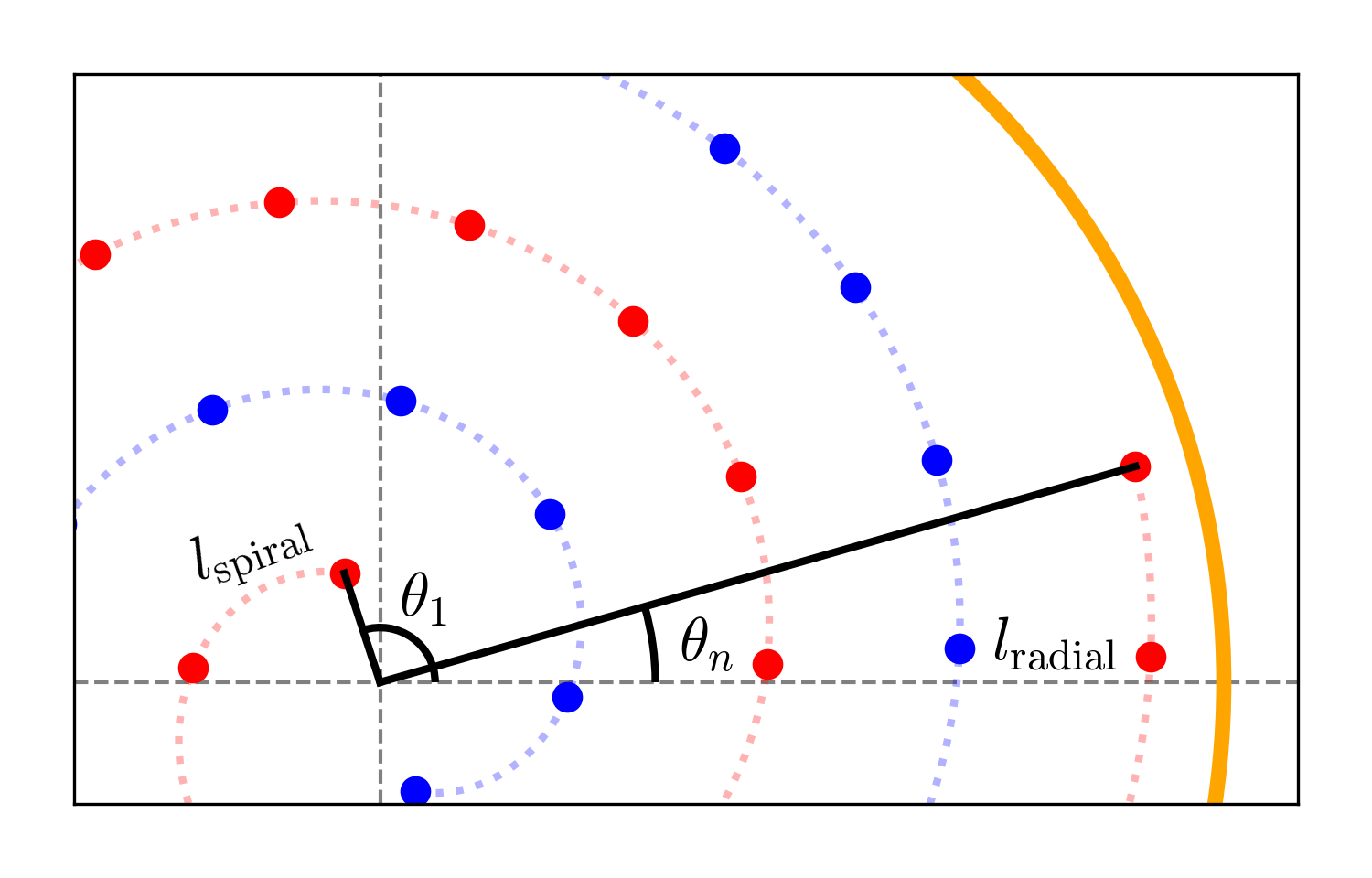}
    \caption{Illustration of the spiral bundle geometry used for wire positioning.
    Red and blue dots represent the locations of metallic wires arranged in a two-arm spiral configuration, with each color corresponding to one spiral arm.
    The initial and final angular positions are denoted by $\theta_1$ and $\theta_n$, respectively.
    The spiral arc length is labeled as $l_{\rm spiral}$, while the radial length is indicated as $l_{\rm radial}$.
    The outermost orange boundary denotes the cavity wall.}
    \label{fig:spiral}
\end{figure}

\begin{figure}
    \centering
    \subfloat[$30^\circ$]{%
        \includegraphics[width=0.32\linewidth]{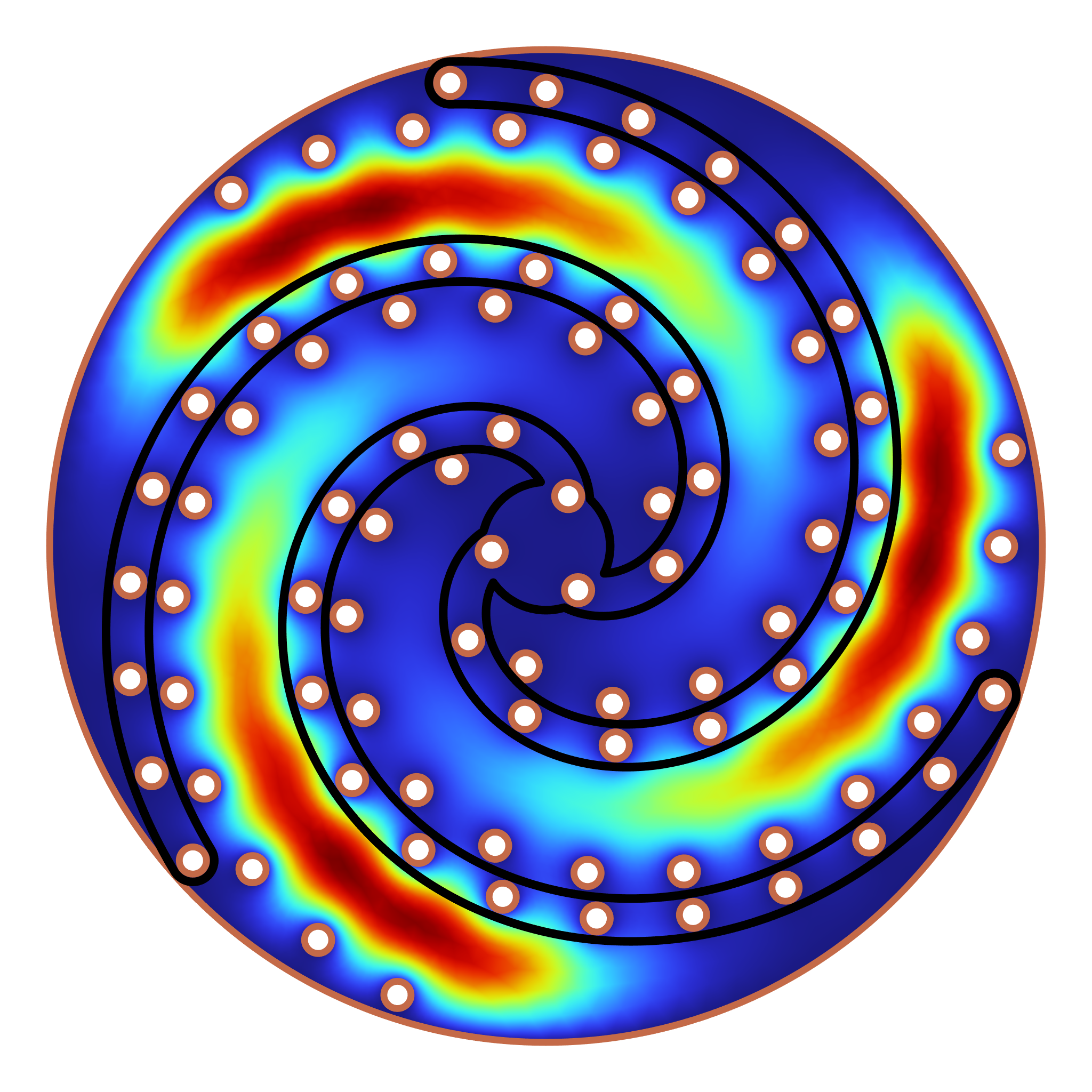}}
    \subfloat[$15^\circ$]{%
        \includegraphics[width=0.32\linewidth]{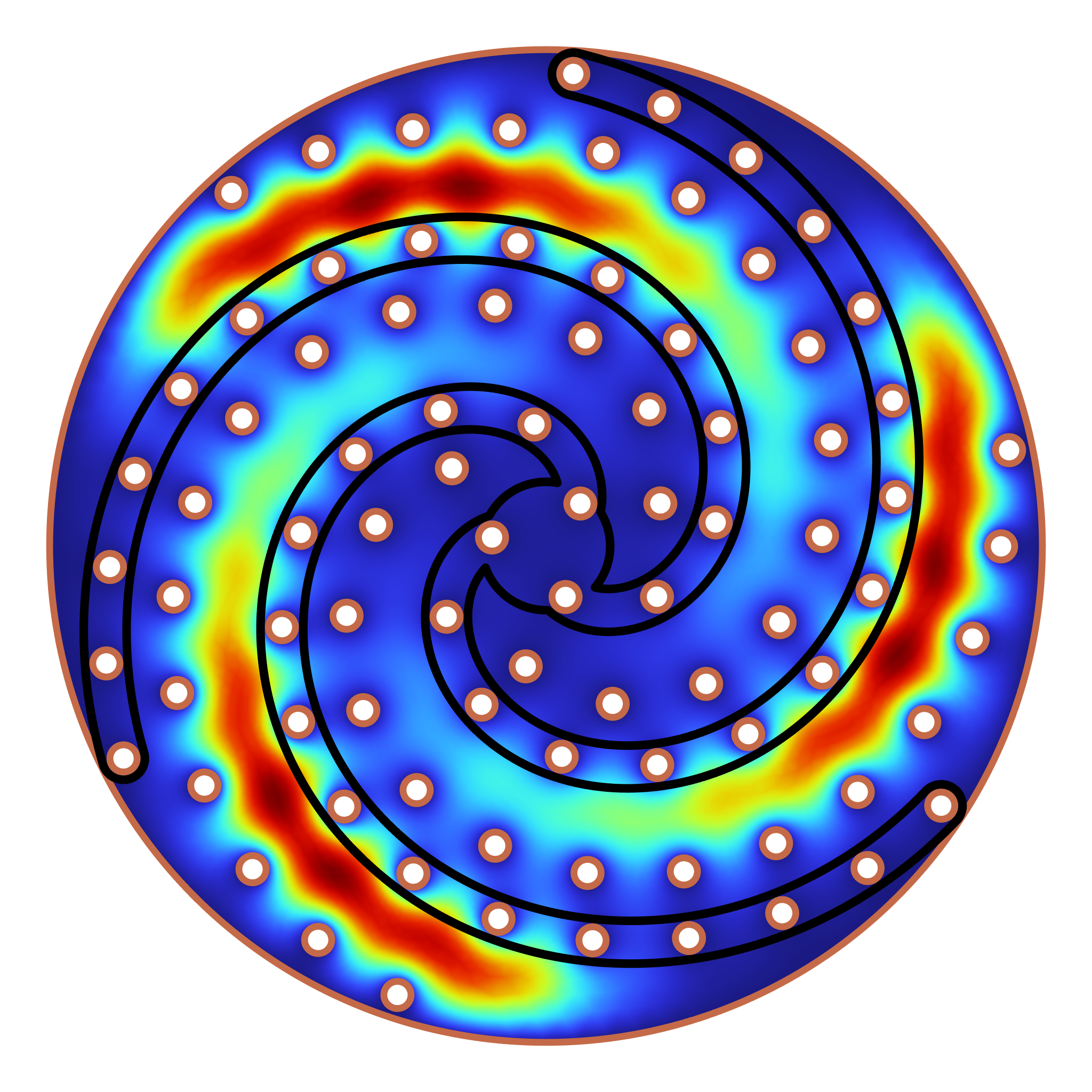}}
    \subfloat[$0^\circ$]{%
        \includegraphics[width=0.32\linewidth]{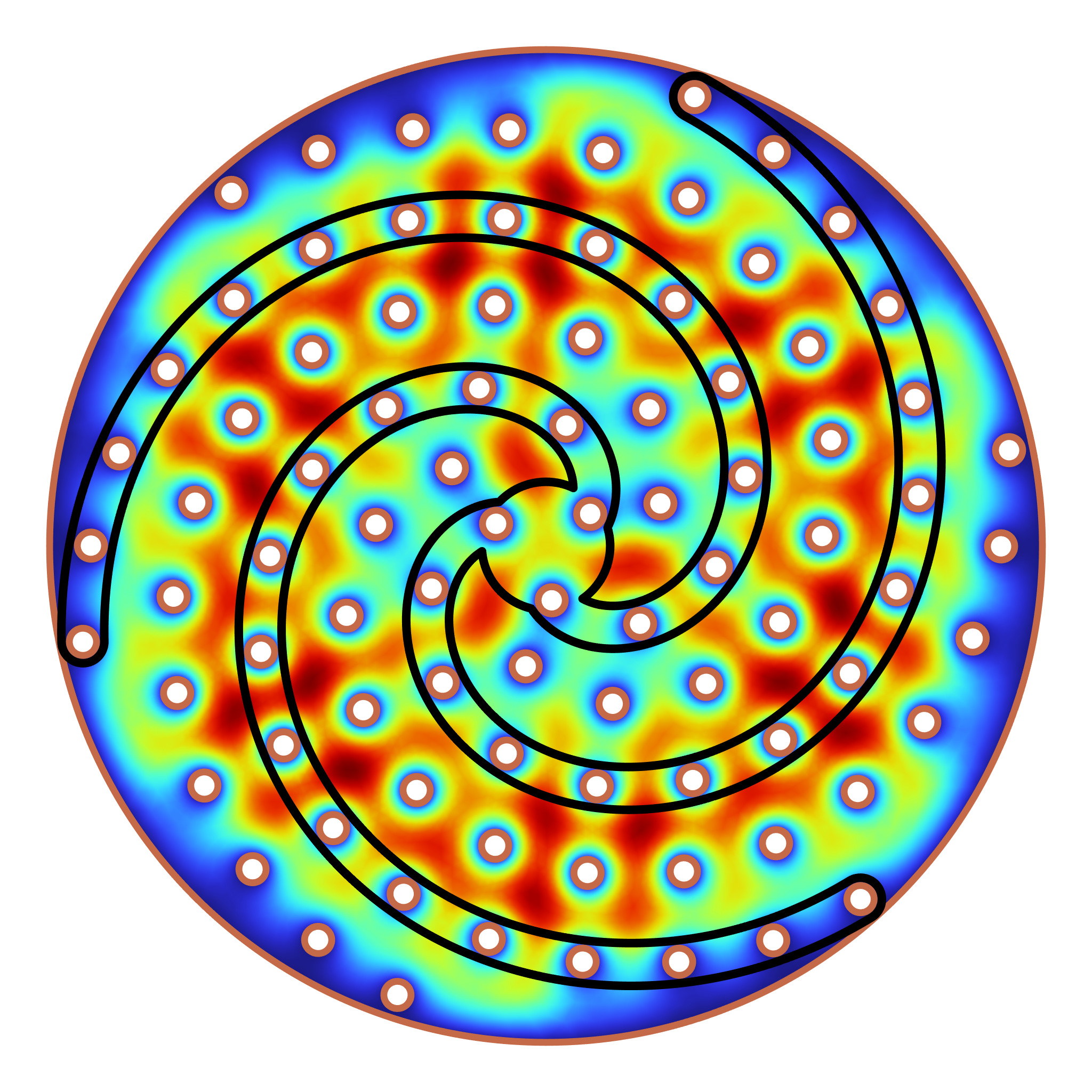}} \\
    \includegraphics[width=\linewidth]{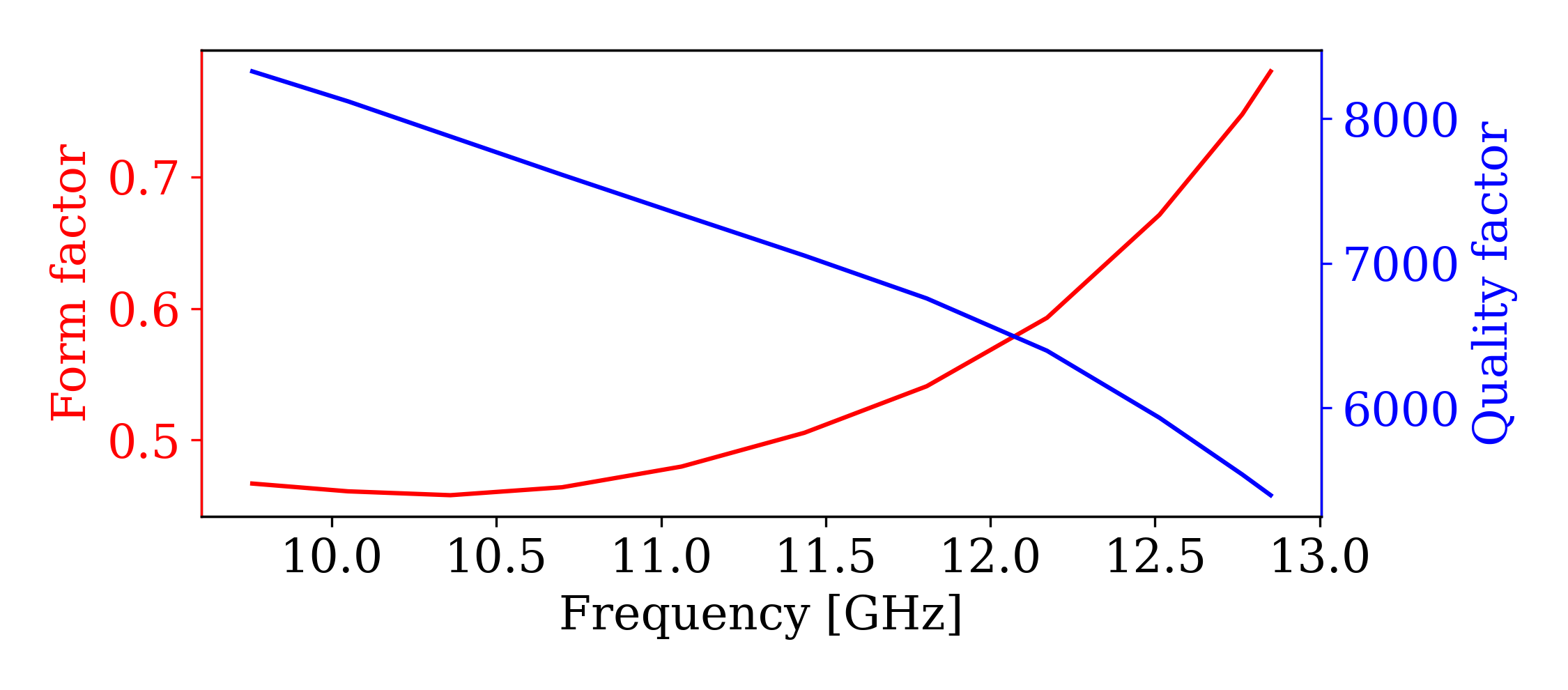}
    \caption{Simulated performance of the spiral tuning mechanism for a six-arm configuration.
    \textbf{Top:} Cross-sectional electric field distributions at various tuning angles (given in the subcaptions), illustrating the evolution of the lowest TM mode across the tuning range. 
    The color scale indicates the electric field strength, with red denoting high intensity and blue denoting low intensity.
    Brown boundaries represent conducting surfaces, and the black spiral lines correspond to the supports holding the rotatable wires.
    \textbf{Bottom:} Simulated form factor (red, left axis) and quality factor for room-temperature copper (blue, right axis) as functions of frequency.}
    \label{fig:spiral_tuning}
\end{figure}

Among various possible geometries, the spiral configuration offers an efficient means to compactly arrange wires within a cylindrical volume, as illustrated in Figure~\ref{fig:spiral}.
This design enables a simple and scalable tuning mechanism.
The spiral arms are divided into two sets of equal size: one fixed and one rotatable.
Rotating one set relative to the other alters the azimuthal spacing between adjacent wires, thereby tuning the cavity's resonant frequency, as shown in Figure~\ref{fig:spiral_tuning}.
This mechanism can be regarded as a polar-coordinate analogue of the linear translation method, employed in the original plasma haloscope designs~\cite{Lawson:PRL:2019}.

Wire positioning along each spiral arm follows a geometric rule designed to maintain approximately uniform spacing in both radial and azimuthal directions.
Consistent spacing in both directions is achieved if the spiral follows a linear polar equation of the form $r = a \theta$ for $\theta \gg 1$, where $a$ is a characteristic length scale.

This spiral configuration is defined by four key parameters: the number of spiral arms $N_s$, the number of wires per spiral arm $n$, the angular position $\theta_1$ of the innermost wire, and the scale factor $a$.
The total number of wires in the cavity is then given by $N_s \times n$.

The remaining parameters are determined so as to yield a uniform distribution of wires.
The radial spacing between adjacent wires is expressed as
\begin{equation}
    l_{\rm radial} \approx \frac{2\pi a}{N_s}.
\end{equation}
To ensure uniform spacing along the spiral, the angular positions of wires are determined by the condition
\begin{equation}\label{eq:spiral_dist}
    l_{\rm spiral} = \int_{\theta_{i}}^{\theta_{i+1}}a\sqrt{1 + \theta^{2}} d\theta = l_{\rm radial}.
\end{equation}
If dielectric materials are present, such as support structures used for spiral bundling in the tuning mechanism, their effect can be accounted for by scaling distances with the square root of the relative permittivity $\varepsilon$, i.e., $l \rightarrow l / \sqrt{\varepsilon}$.

An approximate solution to Equation~\ref{eq:spiral_dist} yields
\begin{equation}\label{eq:spiral_angle}
    \theta_{i} \approx \sqrt{\theta_{1}^{2} + \frac{\theta_{n}^{2} - \theta_{1}^{2}}{n - 1} \cdot (i - 1)},
\end{equation}
where $\theta_i$ is the angular position of the $i$-th wire ($i = 1, 2, \dots, n$), and $\theta_n$ is the angular position of the outermost wire.

The design procedure begins by fixing a cavity radius $r_c$ and wire radius $r_w$.
The number of spiral arms $N_s$ is then chosen based on mechanical considerations.
With a selected value of $n$, the wire density is defined.
The corresponding resonant frequency for uniformly distributed wires is approximately
\begin{equation}
    f \approx \frac{c}{\pi} \sqrt{\frac{N_s \times n}{\text{Cross-sectional area}}},
\end{equation}
where $c$ is the speed of light.
This formula can guide the selection of $n$ to target a desired frequency range.


If no support is present, approximate relations for $\theta_1$ and $\theta_n$ follows below proportionality from the numerical simulations.
\begin{equation}
    \theta_{1} \sim \log n, \quad \theta_{n} \sim \sqrt{n}
\end{equation}
When dielectric supports are included, small corrections to these angles must be introduced to account for the altered boundary conditions.

The tuning mechanism is conceptually simple, as noted earlier.
The underlying principle is illustrated in Figure~\ref{fig:spiral_tuning}.
It depicts a representative example based on a six-arm spiral configuration with the same dimensions as described in the next section.
The wire array is divided into fixed and rotatable sets. 
The rotatable wires are bundled through a spiral-shaped dielectric support, and rotation of this set about the central axis alters the azimuthal spacing between the spiral arms.
This geometric reconfiguration changes the boundary conditions inside the cavity, thereby shifting its resonant frequency.
The spiral configuration gives rise to a galaxy-like electromagnetic field distribution.

The performance of the cavity design is primarily characterized by the form factor and the quality factor.
The expected axion-photon conversion power is proportional to the product of these two quantities, while the scanning speed scales linearly with the quality factor and quadratically with the form factor.

The quality factor characterizes the dissipation properties of the cavity, while the form factor quantifies the overlap between the resonant mode's electric field and the externally applied static magnetic field.
\begin{equation}
    {\rm Form\ factor} = \frac{\left| \int \mathbf{E}\cdot\mathbf{B}_{\rm ext} \, dV_{c} \right|^{2}}{\int \varepsilon \left| \mathbf{E} \right|^{2} dV_{c} \  \int \left| \mathbf{B}_{\rm ext} \right|^{2} \, dV_{c}},
\end{equation}
where $\mathbf{E}$ is the electric field of the cavity mode, $\varepsilon$ is the dielectric constant of the material inside the cavity, and $\mathbf{B}_{\rm ext}$ is the externally applied magnetic field.
In typical haloscope experiments, $\mathbf{B}_{\rm ext}$ is aligned along the axial direction, as the system is placed inside a solenoidal magnet.
In this configuration, the fundamental TM mode provides the highest form factor.

The plot in Figure~\ref{fig:spiral_tuning} shows that the total tuning range exceeds 25\%, with the form factor remaining around 0.5 at low frequencies and increasing at higher frequencies.
The quality factor, estimated using the room-temperature conductivity of copper, decreases at higher frequencies due to increased surface losses, as the fields are uniformly distributed across the conducting wires.

\begin{figure}
    \centering
    \includegraphics[width=\linewidth]{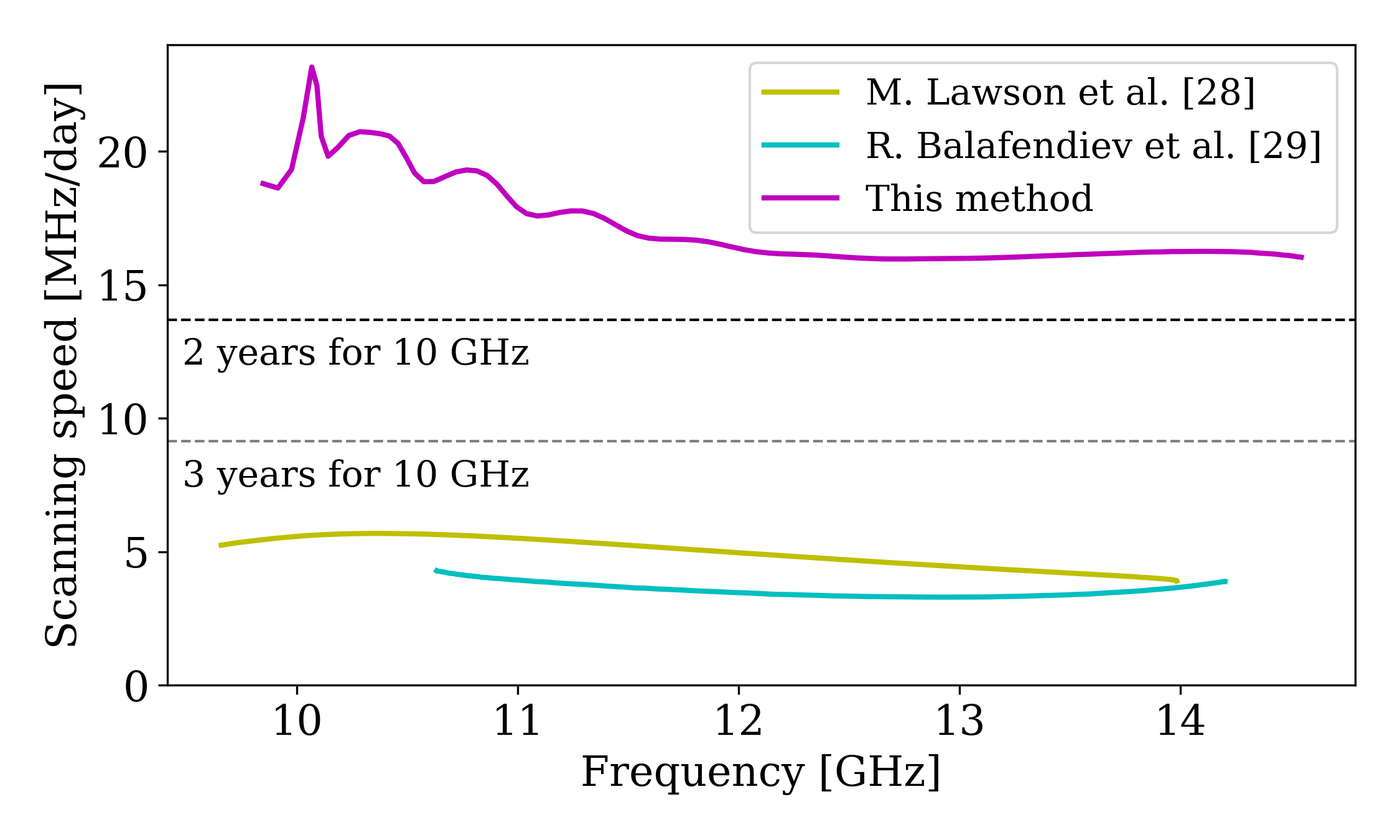}
    \caption{Scanning speed comparison at KSVZ sensitivity for different tuning methods, as indicated in the legend.
    The black and gray dashed lines indicate the scanning rates required to cover a 10\,GHz range within 2 and 3 years, respectively.}
    \label{fig:comparison}
\end{figure}

Figure~\ref{fig:comparison} compares the scanning speeds of two tuning methods for plasma haloscopes~\cite{Lawson:PRL:2019,Balafendiev:arXiv:2025} with the present approach in a two-dimensional study.
The calculation assumes cavities fitting within a 165\,mm inner diameter and 500\,mm inner height, relevant to the ALPHA (Axion Longitudinal Plasma HAloscope) experiment~\cite{Millar:PRD:2023}. 
The spiral method achieves a scanning speed about 3--4 times faster than the alternatives.
This improvement primarily arises from the circular packing geometry of the spiral design, whereas the alternative methods are constrained by rectangular perimeters that significantly reduce the usable volume within a solenoidal magnet.
The wire diameter is chosen to be 0.5\,mm in this study, which enhances the overall frequency tuning range compared to the previous example.
The evaluation further assumes a 9\,T magnetic field, an axion dark matter density of 0.45\,GeV/cm$^{3}$~\cite{Ou:MNRAS:2024,Lim:JCAP:2025}, KSVZ coupling strength, and a standard quantum-limited noise temperature. 
A surface conductivity ten times higher than that of room-temperature copper is also assumed, consistent with typical cryogenic values at microwave frequencies.

\section{\label{sec:meas}Prototype Cavity}

To evaluate the feasibility of the spiral tuning mechanism, we constructed a prototype cavity incorporating six spiral arms.
The cavity body is fabricated from copper with a cylindrical geometry, featuring an inner diameter of 116\,mm and a height of 100\,mm.

The fixed wires are anchored directly to the cavity endcaps, while the rotatable wires are mounted on a spiral dielectric disk.
Two 5-mm-thick disks made of polyether ether ketone (PEEK) were installed at the top and bottom of the cavity to support the rotatable wires while minimizing dielectric loading in the central volume.

The number of wires per arm was set to 15 to achieve a resonant frequency above 10\,GHz.
Each wire has a diameter of 3.175\,mm, corresponding to standard 1/8\,inch copper tubing.

The angular positions of the wires are determined using Equation~\ref{eq:spiral_angle}, with the initial and final positions optimized in \textsc{COMSOL}~\cite{comsol}.
The optimization aims to maximize the frequency of the fundamental TM mode in the unrotated configuration, thereby ensuring a widespread field distribution throughout the cavity volume.

\begin{figure}
    \centering
    \includegraphics[width=\linewidth]{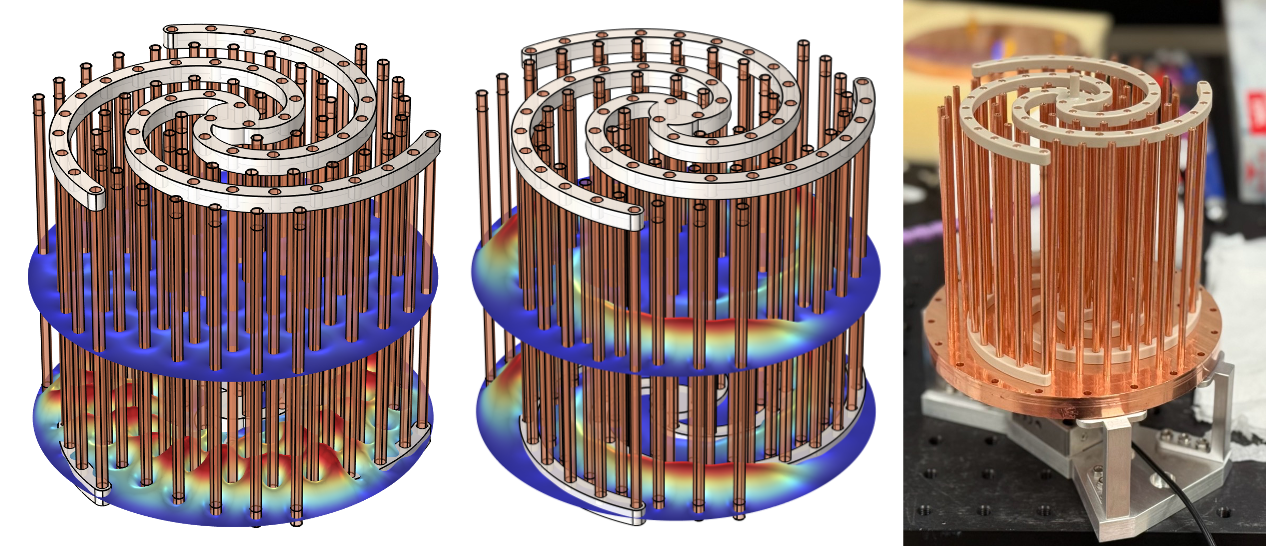}
    \caption{Simulated and experimental images of the six-arm prototype cavity.
    The two panels on the left show the simulated electric field distributions obtained with \textsc{COMSOL} for rotation angles of $0^\circ$ (left) and $30^\circ$ (center).
    Brown vertical rods represent the conducting wires, with one set fixed and the other rotatable.
    The top and bottom spiral-shaped structures are PEEK dielectric supports that hold the rotatable wires in place and define their angular positions.
    The right panel is a photograph of the partially assembled prototype, shown without the outer cylindrical enclosure to reveal the inner spiral structure.
    The spiral support is mechanically coupled to a rotary piezoelectric actuator via its central shaft, which is visible at the bottom of the assembly.}
    \label{fig:assembled}
\end{figure}

Figure~\ref{fig:assembled} shows a photograph of the half-assembled prototype cavity alongside the corresponding three-dimensional simulated electric field distribution.
Each cavity endcap is equipped with three antenna ports to allow flexible coupling configurations.
The spiral support disks incorporate a central shaft extending along the cavity axis, which passes through metal bearings mounted on each endcap to enable smooth, controlled rotation.
One end of the shaft is connected to a piezoelectric rotary actuator, allowing fine tuning of the rotatable spiral wires.

\begin{figure*}
    \centering
    \includegraphics[width=\linewidth]{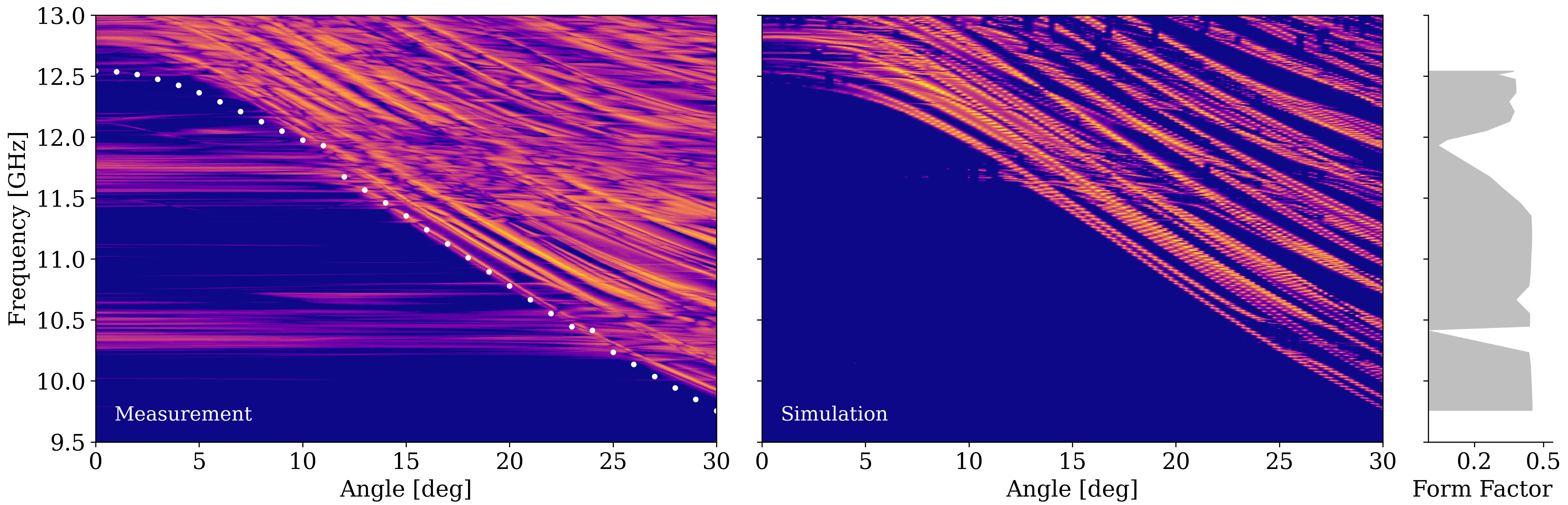}
    \caption{Measured (left) and simulated (center) mode maps of the six-arm prototype cavity as functions of the spiral rotation angle.
    The color scale indicates the transmission amplitude, with brighter colors corresponding to stronger signals.
    The horizontal axis represents the relative rotation angle between the rotatable and fixed spiral bundles.
    White dots in the measured plot indicate the simulated frequencies of the desired mode obtained from an eigenmode study in \textsc{COMSOL}.
    The simulated mode map was generated using \textsc{CST}~\cite{cst}.
    Missing resonances in the simulated map at certain angles arise because \textsc{CST} skipped some solutions during the frequency sweep.
    The gray curve on the right shows the simulated form factor of the mode.}
    \label{fig:modemap}
\end{figure*}

Tuning is achieved by rotating the central shaft.
The cavity resonant modes were characterized through reflection and transmission measurements using two weakly coupled antennas mounted on the endcaps.
As the spiral arms were rotated, transmission spectra were recorded at room temperature, and the resulting mode map is shown in Figure~\ref{fig:modemap} alongside the simulation for comparison.

The measured mode map shows good agreement with both \textsc{COMSOL} and \textsc{CST} results, validating the spiral tuning mechanism as a viable approach for plasma haloscope applications.
Our target mode corresponds to the lowest-frequency tuned mode, consistent with the eigenmode solutions indicated by the white dots on the measured map.
Several transverse electric (TE) modes and strong transverse electromagnetic (TEM) modes mixing appear as nearly horizontal bright lines.
The mixing band appears broadened, as expected from imperfect wire alignment on the supports, machining and assembly errors, and possible support bending.
The number of undesired TE mode crossings can be further reduced by implementing a photonic bandgap structure~\cite{Lewis:arXiv:2024,Goulart:arXiv:2025}.

The achieved tuning range is approximately 25\% in frequency.
Three-dimensional electromagnetic simulations show that the form factor remains around 0.45, with a decrease in the highest-frequency region caused by mode mixing and field localization.
This effect arises from the dielectric supports holding the wires at the top and bottom, as well as their gaps to the endcaps, which break the longitudinal symmetry.
As a result, both the frequency coverage and overall form factor are reduced compared to the two-dimensional study.


\section{\label{sec:discussion}Discussion}

A key advantage of the spiral-arm geometry is its ability to accommodate a high density of conducting wires with minimal mechanical complexity.
Conventional metamaterial cavities typically require multiple independent tuning shafts, since the spacing between wires must be individually controlled in each unit cell~\cite{Balafendiev:arXiv:2025,Goulart:arXiv:2025}.
In contrast, the spiral-bundling approach enables collective tuning of the wires through a single rotational degree of freedom, while maintaining a relatively stable form factor across the tuning range.

Enhanced mode mixing arises from longitudinal symmetry breaking introduced by the spiral support disks.
These dielectric supports could be extended to span the full height of the cavity or replaced with a conducting support concealed from the cavity interior, thereby suppressing arm bending and improving axial symmetry.
Such a modification is expected to reduce mode hybridization and localization while enhancing the overall electromagnetic performance.


The quality factor expected in this design can be significantly improved through the use of superconducting cavity techniques~\cite{Alesini:PRD:2019,Ahn:PRA:2022,Ahyoune:JHEP:2025}.
The dominant source of loss arises from the large number of conducting surfaces introduced by the wire array.
Replacing the copper wires with superconducting ones is expected to substantially increase the quality factor by suppressing ohmic losses.

The spiral geometry presented here is well-suited for high-mass axion search experiments using plasma haloscope techniques operating above the 10\,GHz frequency range.
Its scalability and tunability make it a promising cavity design for the ALPHA experiment, providing the technology required to probe the QCD axion in the 10--50\,GHz band with quantum-noise-limited sensitivity.

\section{\label{sec:conclusion}Conclusion}

We have proposed and demonstrated a novel spiral-based tuning mechanism for plasma haloscope cavities.
This approach enables dense packing of metallic wires within a cylindrical volume while permitting mechanically simple tuning through a single rotational axis, independent of the number of wires.
Two-dimensional electromagnetic simulations show a tuning range exceeding 25\%, with the form factor remaining approximately 0.5 or higher across the frequency coverage.
This method achieves scanning rates several times faster than traditional plasma haloscope tuning approaches, primarily due to the circular perimeter.
The feasibility of the concept was demonstrated with a six-arm prototype cavity, showing good agreement with simulations in its frequency tuning characteristics.
This cavity design offers a promising route toward scalable and tunable high-frequency structures for future plasma haloscope experiments, including ALPHA.

\begin{acknowledgments}
This research is supported by the Knut and Alice Wallenberg Foundation, Olle Engkvists Foundation, and the Swedish Research Council (VR) under Dnr 2019-02337 ``Detecting Axion Dark Matter In The Sky And In The Lab'' (AxionDM).
The authors gratefully acknowledge valuable discussions within the ALPHA Collaboration.

\end{acknowledgments}

\bibliographystyle{apsrev4-2}
\bibliography{main}

\end{document}